# Contextual Information Enhanced Convolutional Neural Networks for Retinal Vessel Segmentation in Color Fundus Images

Muyi Sun[a,b,1], Kaiqi Li[b,1], Xingqun Qi[a], Hao Dang[a,b], Guanhong Zhang[c,∗]

*[a]CRIPAC, CASIA, Beijing 100876, China*
*[b]School of Automation, Beijing University of Posts and Telecommunications, Beijing 100876, China*
*[c]School of Computer Science, Beijing University of Posts and Telecommunications, Beijing 100876, China*

**Abstract**

Accurate retinal vessel segmentation is a challenging problem in color fundus image analysis. An automatic retinal vessel segmentation system can effectively facilitate clinical diagnosis and ophthalmological research. Technically, this problem suffers from various degrees of vessel thickness, perception of details, and contextual feature fusion. For addressing these challenges, a deep learning based method has been proposed and several customized modules have been integrated into the well-known encoder-decoder architecture U-net, which is mainly employed in medical image segmentation. Structurally, cascaded dilated convolutional modules have been integrated into the intermediate layers, for obtaining larger receptive field and generating denser encoded feature maps. Also, the advantages of the pyramid module with spatial continuity have been taken, for multi-thickness perception, detail refinement, and contextual feature fusion. Additionally, the effectiveness of different normalization approaches has been discussed in network training for different datasets with specific properties. Experimentally, sufficient comparative experiments have been enforced on three retinal vessel segmentation datasets, DRIVE, CHASEDB1, and the unhealthy dataset STARE. As a result, the proposed method outperforms the work of predecessors and achieves state-of-the-art performance in Sensitivity/Recall, F1-score and MCC.

## 1. Introduction

The eye is one of the most important organs for humans to receive and perceive external information. In the total amount of information received by ordinary people, the information received by the eyes can account for more than 80%. The retina of the fundus is one of the most important parts of the eye [1]. In the field of detecting diseases of eye, cardiovascular and cerebrovascular, such as glaucoma, diabetes, hypertension, and atherosclerosis, doctors can effectively screen and judge disease types and causes by examining and analyzing the morphology or structure of retinal blood vessels in the fundus [2]. Therefore, fundus retinal examination is an important part of the eye examination. In this situation, the most important portion of the examination is to extract the morphology and structure of retinal blood vessels. Color fundus imaging is a widely used non-invasive way for directly imaging of the fundus microvascular system in clinical practice [3]. In the traditional medical process, it is necessary to rely on experienced experts to manually segment the retinal blood vessel area. However, the blood vessels in the retinal images are distributed densely and irregularly [4]. There are a large number of small blood vessels which have low contrast and could be easily confused with the background. These problems lead the manual segmentation of retinal vessels becoming a cumbersome, inefficient, and inconsistent task. Therefore, an automatically accurate segmentation method for retinal vessels is indispensable.

In the previous research of retinal vessel segmentation, there are three types of methods, traditional digital image processing segmentation methods, machine learning models based on supervised learning and feature engineering, and deep learning based approaches.

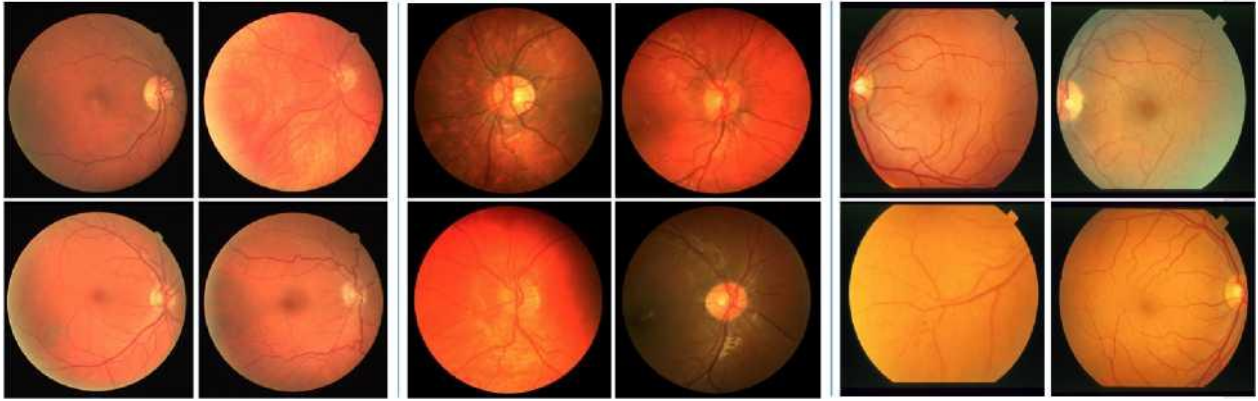

Figure 1: Several samples randomly sampled from these two datasets DRIVE and CHASEDB1 separated by the blue line. **Left** : Four samples from DRIVE dataset. They are controlled for similar brightness, contrast, and so on. **middle** : Four samples from CHASE_DB1 dataset. The properties of these images are totally different. **right** : Four samples from STARE dataset. The properties of these images are totally different. The images in CHASE_DB1 dataset are more various than images in DRIVE. The images in unhealthy dataset STARE are various and several in special color.

In the field of digital image processing segmentation methods, Katz and Nelson [5] have firstly introduced a Gaussian function into the blood vessel segmentation task on the fundus retina image in 1989. They have designed a two-dimension matched filters for vessel detection. The matched filter response is improved by Ant colony algorithm, which is proposed in [51]. Li et al. [52] propose that multiscale matched filter is a better approach compared with single scale filter. Bob et al. [53] employs Zero mean Gaussian matched filter with first order derivative. The approaches based on matched filters are fully discussed by Sreejini and Govindan [50], in which particle swarm optimization is used to generate a optimized set of multiscale Gaussian matched filters. Jiang and Mojon [6] have proposed an adaptive local thresholding method based on verification-based multi-threshold probing. Elisa and Renzo [7] have used line operators and support vector classification method for extracting the target areas. Yang et al. [8] have combined fuzzy C-means clustering and digital morphology to extract vascular regions. However, the robustness of these methods is relatively poor. These methods need priori knowledge to design handcrafted features for specific datasets, which impair the generalization ability of these approaches on new datasets.

For addressing the problem about robustness above mentioned, several machine learning methods have been proposed. Staal and Joes [9] have classified each pixel using the K- nearest neighbor method for recognizing pixels of the class of vessel. Soares et al. [10] have utilized a two-dimensional Gabor filter to extract the features of the retinal image and then used a Bayesian classifier to classify the target and background categories. Osareh et al. [11] have employed multi-scale filtering and principal component analysis to extract features and reduce the feature vectors and then classified these vectors with the integrated model of Gaussian mixture model and support vector machine. Sunder et al. [12] have proposed a framework using hybrid feature set and hierarchical classification for retinal vessel segmentation.

With the rapid development of deep learning in medical image analysis, deep learning based methods have overcome the feature extraction based machine learning methods and become the mainstream [13]. Research about fundus retinal vessel segmentation in color fundus images with deep neural networks has attracted more attention in recent years. Liskowski [14] has proposed a patch-based shallower convolutional neural network to classify the category of pixels in the center of the patch. Melinscak [15] and Li [16] have applied different convolutional neural network structures which developed in the field of image recognition, to retinal vessel segmentation tasks. In 2015, FCN-based neural networks have rapidly become the main methods of various segmentation tasks in the field of computer vision with the proposal of Fully Convolutional Network (FCN) [17]. Orlando et al. [18] have employed a discriminatively trained fully connected conditional random field model for the retinal vessel segmentation task. Alom et al. [19] have applied U-Net to the retinal vessel segmentation task and proposed a cyclic U-Net in combination with the idea of a recurrent neural network, further improving the accuracy of segmentation. Zhuang et al. [20] have designed LadderNet which could use stacked U-Net to hybridize the application of multi-scale features. Yan et al. [21] have proposed joint segment-level and pixel-wise losses for the segmentation of thin vessels. And in 2019, Tim et al. [22] have proposed an effective and efficient lightweight network for ARM-based equipment. These methods have achieved state-of-the-art performance on the datasets of DRIVE, CHASEDB1 and STARE. However, despite these research has achieved significant progress, there are still many challenges in the field of the retinal vessel segmentation, such as the recognition of various degrees of vessel thickness, the perception of details, and the effectiveness of contextual feature fusion [25].

Motivated by the above observations, in this paper, a Contextual Information Enhanced dilated convolutional networks (CIEU-Net) based on U-Net has been proposed. For accurate segmentation, cascaded dilated convolutional modules with a multi-grid strategy have been integrated into the intermediate layers [38]. With the atrous operations, the dilated convolutional modules could obtain larger receptive fields and fuse contextual features. Meanwhile, the advantages of pyramid modules have been taken for fusing multi-scale features and enhancing the details, especially for the small blood vessels which have low contrast and can be easily confused with the background. Specifically, the pyramid module with spatial continuity has been used due to the consecutive property of vessels. Meanwhile, the effectiveness of different normalization approaches has been discussed in network training on the datasets with uniform or random properties such as brightness, contrast, saturation. Though in previous methods, researchers have used the Batch Normalization (BN) method uniformly. As a result, the employments of different normalization approaches have been summarized for different datasets with specific properties.

Contributions of this paper can be summarized as:

1) . A Contextual Information Enhanced U-net — CIEU-Net has been proposed with dilated convolutional modules for accurate retinal vessel segmentation in color fundus images.

2) . The cascaded dilated modules and the pyramid module with spatial-continuity have been

integrated in our network for detail and multi-scale perception.

3) . The effectiveness of different normalization approaches for different datasets with specific properties has been discovered, which is usually ignored in the field of medical image analysis.

4) . Sufficient comparative experiments have been enforced on the datasets DRIVE, CHASEDB1 and the unhealthy dataset STARE, and the results have achieved state-of-the- art performance.

## 2. Involved techniques

In this section, the related work in medical image analysis and natural image semantic segmentation is introduced. In recent years, more and more research is concentrated on the combination of medical image characteristics and semantic segmentation methods in computer vision [13]. Following this trend, in this paper, the classical methods of medical image segmentation have been inherited and the novel modules from the field of image semantic segmentation have been integrated.

*2.1. Encoder-Decoder architecture in medical image analysis.*

Encoder-Decoder architecture is one of the most widely used fully convolutional network in image segmentation. And in medical image segmentation tasks, U-net [26] is one famous sample of encoder-decoder architecture which is firstly designed for cell segmentation in electron microscope images. U-Net employs an encoder-decoder structure with skip-connections between the feature maps with the same output stride, as illustrated in Fig. 2(a). Through these connections, U-Net could use shallow features with rich details but low semantic-level to optimize and correct the deep features with higher semantic-level.

In the field of medical image segmentation, U-Net is recognized as a basic contrastive method which has been used, compared, or improved in nearly all aspects such as organs segmentation in CT [27], tissue/nucleus segmentation in digital pathology image [28], pulmonary nodule detection in MRI [29] and cardiac blood vessel segmentation in angiographic imaging [30]. In this research, the backbone architecture has been designed based on the idea of U-Net with 47 convolutional layers.

*2.2. contextual modules in semantic segmentation.*

In the field of semantic segmentation, contextual information is indispensable for image scenes parsing [31]. Obtaining global information with the correlation of different regions or pixels is important for image understanding. In the experiments of this research, two customized contextual information enhancing modules, cascaded dilated modules and pyramid module with spatial continuity have been utilized.

**Dilated convolution.** Deep convolutional neural networks have been proved to be effective in semantic segmentation task [32]. In [33], fully convolutional neural networks have been employed to segment retina vessel. In [34] and [35], full-resouluton networks have been used to settle the high-resolution images. However, the pooling or striding operations which are repeatedly used in these frameworks significantly reduce the resolution of the final feature maps. Obviously, this is harmful for dense prediction. Dilated convolution, or atrous convolution is a corresponding method to solve this problem which has been firstly implemented in a dyadic wavelet transform method [36] in the field of signal processing. Chen and Papandreou [37] have firstly employed atrous convolution in semantic segmentation to replace these resolution-reducing layers. In [38], a similar approach has been adopted for increasing receptive fields and preserving the resolution of feature maps simultaneously.

**Cascaded dilated modules.** For enhancing the functionality of dilated convolution and obtaining larger receptive field, stacked dilated convolutions (dilated module) have been proposed in [39]. Chen and Papandreou [40] have integrated the cascaded dilated modules in DeeplabV2, and in [41], they have further discussed the multi-grid strategy in cascaded dilated modules for gaining better performance. Wang and Chen [42] have improved this module and explained the gridding problem specifically corresponding to the multi-grid strategy. In this paper, three modules with multi-grid strategy have been utilized in the backbone architecture, which is illustrated in Fig. 2(b).

***Spatial pyramid module.*** Pyramid structure is a classical technique in image processing [43]. With the development of deep learning, spatial pyramid pooling module has been firstly proposed in SPP-Net [44] for multi-scale object detection. PSPNet [45] has used a similar strategy to serve the multi-scale feature extraction. Atrous Spatial Pyramid Pooling [40] is an advanced structure based on atrous convolution. This module has been employed in DeepLab series [37], [40], [41] to capture more comprehensive context information. In the network of this paper, the module as PSPNet has been used, which is relatively continuous in spatial, corresponding to the continuous property of vessels.

*2.3. Normalization of convolution.*

In order to increase training stability, reduce the possibility of gradient disappearance or gradient explosion, and improve the training efficiency, normalization of convolutional layers has made a huge contribution.

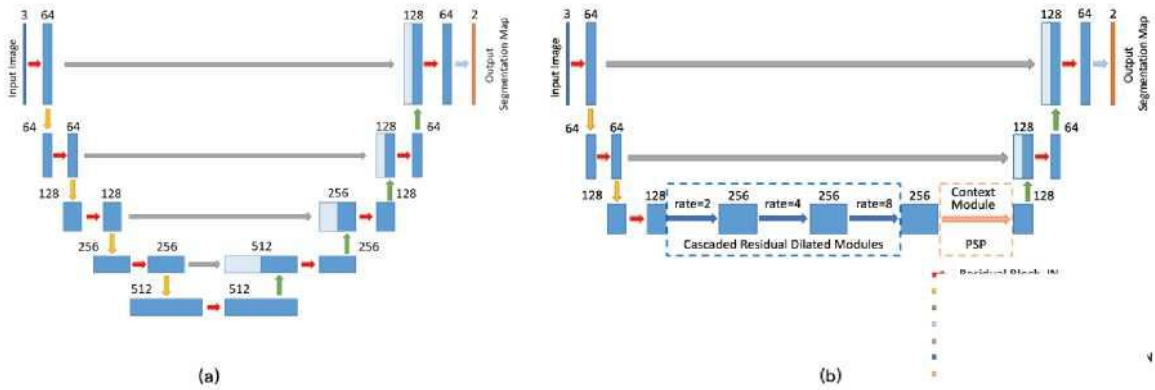

Figure 2: Two architectures used in our experiments. **(a).** The baseline U-Net architecture. **(b).** The proposed architecture which integrated with cascaded residual dilated modules and PSP context module. The multi-grid strategy has been applied on each dilated convolution module to change the dilation rates of dilated convolution blocks to (1,2,1), (2,4,2), (4,8,4)

Batch normalization (BN) [46] is one of the earliest widely used inter-layer normalization strategy of convolutional neural networks, aiming at solving the covariance drifting problem of feature distribution between activation layers. Group normalization (GN) [47] is a simplified version of BN. The huge computational requirements of BN make this normalization strategy unsuitable for use in lab environments, embedded systems, or mobile devices where computing resources are limited. The motivation of GN is to design a model which could be effectively normalized when the batch size is small and the computing resources are limited. Instance normalization (IN) [48] only standardizes the features of a single sample, which is suitable for domain transfer learning. However, in the field of medical image analysis, researchers have shown more attention to the network architectures and ignored the role of normalization. In this paper, these different normalization strategies have been compared on different datasets. The results have shown that for different types of datasets, this type of segmentation problem could be settled with the idea of batch learning or style transfer/domain migration [49].

## 3. Datasets

In this research, three open source datasets have been used in our experiments, DRIVE (Digital Retinal Images for Vessel Extraction) dataset, CHASE_DB1 (Child Heart and Health Study in England) dataset and STARE (STructured Analysis of the Retina) dataset. These datasets are composed of two parts: color fundus images, and the expertly labeled segmentation results corresponding to these images one by one.

The DRIVE dataset and the CHASE_DB1 dataset are the most commonly used evaluation datasets. Most of the previous work based on the deep learning technology routeline has been carried out on these data sets. These datasets can be used to compare model effects with broader methods.

The STARE dataset is a retina vessel segmentation dataset containing some pathological abnormalities. This dataset has the ability to evaluate the effect of the model on abnormal fundus images.

In DRIVE dataset, there are 40 samples which are divided into a training set and a testing set, each containing 20 different color fundus images with resolution 565x548. The CHASEDB1 dataset is a subset of color retinal images extracted from the UK Children's Heart and Health Research Center database, containing 28 color fundus images, each of them has 999x960 pixels in size. Compared with DRIVE, the background brightness of the CHASE_DB1 data set is relatively variable, and the contrast between the background and the blood vessel target is relatively low. STARE is an unhealthy dataset which includes 40 color retinal images, among which 10 images contains pathology sign. The image has a resolution of 605x700 pixels for all images. Manual annotations labeled by the first expert have been used to train and evaluate our model. The same division strategy (tenfold cross validation) has been conducted as previous methods, because the training and testing sets are not specified for the dataset [21]. Similar to CHASE_DB1, images in the unhealthy dataset STARE is also various. The detailed information of these datasets is shown in Table.I. Some samples from these three datasets are illustrated in Fig. 1. In the experiments, patch-based images have been employed as the network input. The original images have been randomly cropped into 64x 64 patches. So in each experiment, the total patches used equals to (iteration = 30,000) x (batch size = 64) = 1,920,000. Basic data augmentation strategies have also been applied such as flipping, rotation, shifting.

## 4. The proposed approach

In this section, U-Net is firstly introduced, which is a widely used deep segmentation framework in medical image analysis. Then some prominent modules developed from the field of semantic segmentation are introduced, which conducted in the experiments. Also, the normalization options are discussed, which should be considered together with the specific properties of different retinal datasets. Finally, the complete network structure is described, which could take full advantages of the knowledge in the two domains.

*4.1 Fully Convolutional Network and U-Net*

Fully Convolutional Network has been proved that it is the most optimal approach in image segmentation task to date [17]. Compared with vanilla convolutional neural networks designed for the

| Dataset | Total | Training | Testing | Resolution | Patch Size | Number of Patches |
|---|---|---|---|---|---|---|
| DRIVE | 60 | 40 | 20 | 565x548 | 64x64 | 1920000 |
| CHASEDB1 | 28 | 8 | 20 | 999x960 | 64x64 | 1920000 |
| STARE | 40 | 36 | 4 | 605x700 | 64x64 | 1920000 |

classification task, FCN removes the last fully connected layer or global average pooling layer to generate dense prediction masks for segmentation mission. However, due to the downsampling operations such as max pooling or strided convolution, the network without fully connected layer can only generate highly-semantic but downsampled prediction results, of which the resolution is limited and determined by the downsampling stride of the backbone network. Features would be more semantic during the hierarchical network progressing, which means higher accuracy in classification, but more degenerative about the segmentation details.

Table 1: The detailed information of these three datasets. In STARE dataset, ten-fold cross validation method has been used for final evaluation.

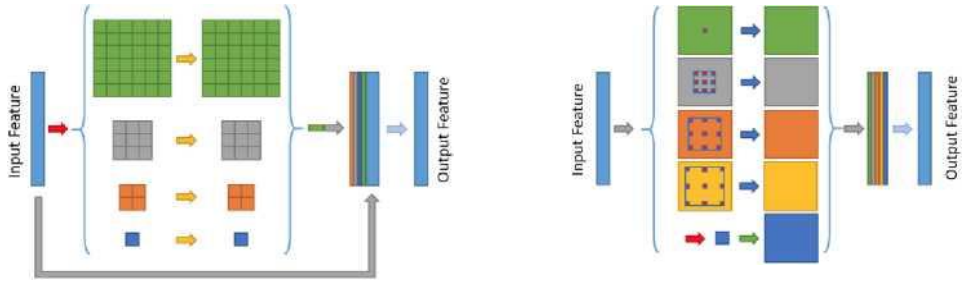

Figure 3: Two contextual information Enhancing modules used in our experiments. **(a).** The PSP module which fuses the global information and pyramid pooling information with spatial-continuity. **(b).** The ASPP module which fuses contextual information by atous convolutions with different dilation rates and global average pooling. Compare these two modules, PSP module preserves more information about spatial continuity and has smaller computational cost.

To alleviate this problem, U-Net integrates the high-resolution and but not highly semantic features with highly-semantic but low-resolution feature maps to generate high- resolution prediction results by skip-connections and layer-by-layer decoder structure. As with most FCN segmentation approaches, U-Net uses autoencoder as basic architecture. U-Net aggregate equivalent feature maps between encoding/downsampling progress and de- coding/upsampling operations with skip-connections. Each feature map generated by upsampling operation would be concated with feature maps of the same resolution in encoding progress at channel dimension, followed by convolutional and nonlinear layers to compress the number of channels and abstract semantic information. The process is repeated layer- by-layer to decode the downsampled features and predict dense segmentation masks which have the same resolution with input images. The layer-by-layer decoder used by U-net and its variants can effectively refine details of low-resolution segmentation mask predicted by encoder network.

In this task, U-Net has been used as the baseline model. The illustration of U-Net is shown in Fig. 2(a).

*4.2. Cascaded dilated Convolution*

As described above, FCN could naturally downsample the intermediate features and prediction masks, as a result, the final segmentation masks are degenerative in details and edges. For the purpose of generating segmentation masks with both accuracy and details, U-Net uses a layer-by-layer decoder to handle this problem. Meanwhile, there is another design route in the semantic segmentation area, which called backbone-based methods. The divergence between these methods and vanilla FCN is that backbone-based methods remove traditional downsampling operations such as max pooling or strided convolution. However, with the extending of depth, the receptive fields of convolutional networks need to be enlarged synchronously to fuse more spacial and semantic information, which is the function of downsampling operations. This is the motivation of dilated convolution which can explicitly enlarge the receptive fields without downsampling. The dilated convolution can be described as:

$$y[i] = x[i+r \cdot a]/[a]. \qquad (i)$$

where x[i] is the input feature maps, y[i] is the output feature maps, *i* means the convolutional center position on feature maps, /[a] denotes a filter of width a with dilated rate *r* which is used to sample the input feature map x[i]. A semantic segmentation network with dilated convolution convolves the input *x* with upsampled filters constructed by inserting r — 1 zeros between each pixel in the convolutional kernels. This operation could be applied iteratively to replace all downsampling operations for maintaining the same resolution as the input image. In practice, the dilated convolution is usually applied properly on downsampled feature maps to find the trade-off between accuracy and computing efficiency. Inspired by [32], the dilated convolutions and ResBlocks have been integrated to build dilated convolution module. Then this module is cascaded to replace the encoder-decoder part under 1/4 resolution of our baseline U-Net.

The cascaded dilated modules could hold the same receptive field size with U-Net to abstract semantic information. Meanwhile, they could generate 1/4 resolution predictive masks compared with 1/16 resolution of encoded outputs without decoder.

As [38] has shown, the multi-grid strategy is essential for dilated convolution to avert gridding problem, which means the convolution kernel convolves in a checkerboard fashion and can only access limited pixels in the receptive field. In each dilated convolutional block, the multi-grid rates have been set to (1,2,1), multiplied with the base dilated rate of the block. The illustration of the multi-grid strategy is shown in Fig. 4, Line. 2.

*4.3. Contextual Information Enhancing Module*

In semantic segmentation task, the segmentation accuracy is extremely subject to context information. Because of the complexity and variance in different semantic classes and objects, recent semantic segmentation networks generally use contextual information enhancing modules to explicitly fuse context information into encoded features for constraining the predicted semantic classes and objects. Most of these modules utilize multi-scale pyramid structures to take full advantages of the global context and partial context. In this paper, two representative contextual information enhancing modules have been chosen to examine if the contextual information is effective in retinal vessel segmentation task as it functions in semantic segmentation task, namely Pyramid Spatial Pooling module (PSP Module) and Atrous Spatial Pyramid Pooling module (ASPP module). In PSP module, the input feature maps are downsampled parallelly by pooling operation, and then directly upsampled and concatenated in channel dimension. Finally, a convolution layer is used to compress the number of channels. With multi-scale pooling operation, the module could refine the integral structure of the vessel region by more holistic and more contextual information. As a result, the final predicted vessel regions achieve better holistic morphology. The architecture of PSP module is shown in Fig. 3(a), in which the pooling kernel sizes are 1x1, 2x2, 3x3 and 6x6 respectively. On the other hand, ASPP module fuses the multi-scale context information by parallelled dilated convolutions. With pyramid dilated rates, ASPP module can directly apply different receptive fields on each parallelled stream of dilated convolution, and merge multi-scale features synchronously, as shown in Fig. 3(b). This module could also utilize more global and contextual features to improve segmentation accuracy.

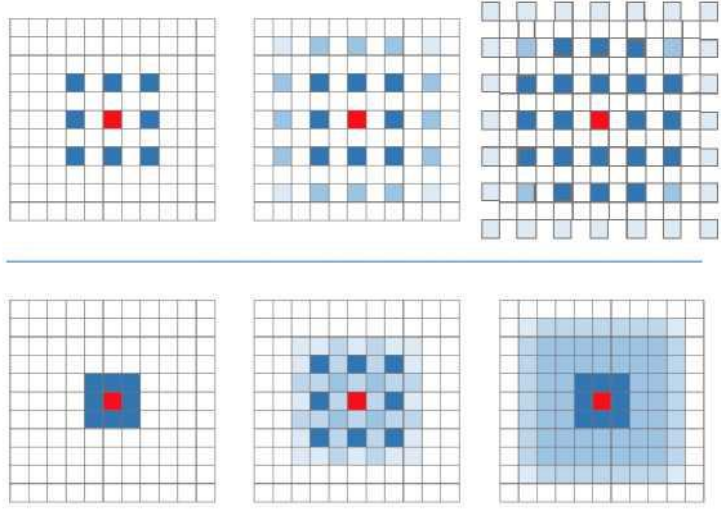

Figure 4: The receptive fields of cascaded dilated modules. **Line 1.** The ordinary cascaded dilated modules with dilation rates = (2,2,2). The illustration shows that the receptive field is incontinuous. This may cause gridding problem. **Line 2.** The cascaded dilated modules with multi-gird strategy in which dilation rates =(1,2,1). The receptive field is enlarged with continuous region.

These selected contextual information enhancing modules have been attached after the cascaded dilated convolutional modules and before the 4 times upsampling layer-by-layer decoder. The contextual information enhancing module is operated on 1/4 resolution feature maps. Then the output of the module is decoded and upsampled to predict final segmentation results, as shown in Fig. 2(b). Based on our experiments, these two contextual information enhancing modules have shown different influence on the final vessel segmentation masks. PSP module tends to force the network generating more aggregated and continuous masks, and ASPP module could rather predict more detailed segmentation results but not as aggregated as the PSP module. To the best of observation, human experts tend to provide a connective domain of vessel regions, hence the PSP module has been chosen in the overall framework which could also produce better segmentation accuracy.

*4.4. Normalization Options*

Normalization operation is an essential component of the convolutional neural network and can dramatically impact the final performance of the network. Batch Normalization (BN) is the most widely used method in classification and semantic segmentation tasks. BN is aimed to avoid the covariance drifting problem of feature distribution between convolution layers. BN computes means and covariances and normalizes feature maps across batch samples sampled from the whole dataset randomly. This means BN can use the information of the whole dataset to correct the distribution of intermediate features and overcome the variant properties of images in dataset such as illumination and rotation degree. It is beneficial for accurate classification mission, as well as pixel-wise classification, namely semantic segmentation. The ideal BN approach should capture the distribution of the whole dataset so the samples number of each batch should be considerably large. This status could result in enormous computation cost of modern convolutional neural networks. Group normalization (GN) is designed to alleviate this problem by a trade-off between computational cost and prediction accuracy. GN computes means and covariances only across feature channel groups on one single sample, not the whole batch samples. Without the variance information of dataset, the classification accuracy is usually reduced as well as the computation cost.

However, in the contrastive experiments, the models with GN have shown better results than BN contrasting on DRIVE dataset, which is nontrivial. Firstly, the differences between BN and GN have been compared. Then Instance Normalization (IN) has been applied on our models. IN is an approach that widely used in style transfer task and a crucial component for this task as a universal consent [49]. IN computes means and covariances of only one feature channel on one single sample, which is equivalent to an extreme GN operation that makes the number of groups equal to channels. Compared with BN, IN only cares about the information of input sample itself, forcing the network to preserve some information such as structure and shape and change other information, for instance, texture and illumination. The experiment results have denoted that IN baseline model gives the best performance of these normalization options on DRIVE dataset. The vessel region segmentation on DRIVE dataset is more similar to a style transfer mission but not a trivial pixel-wise classification task. On the contrary, the BN models have shown better accuracy than IN on CHASEDB1 dataset. The results have shown that normalization option is central to vessel segmentation task, which should be taken into account with data distribution jointly. In the case that the variances of retina images such as illumination condition and rotation degree could be controlled, IN would be a more appropriate choice. Otherwise, BN should be applied to overcome intra-dataset variance.

*4.5. Overall architecture*

Our overall model is shown in Fig. 2(b). A 47-layer residual connected [50] U-net is used as the baseline. For accurate segmentation, the cascaded dilated modules with multi-grid strategy and pyramid module with spatial-continuity have been integrated. Furthermore, the normalization methods for different datasets have been discussed. The details can be referred to as the illustration in Fig. 2(b).

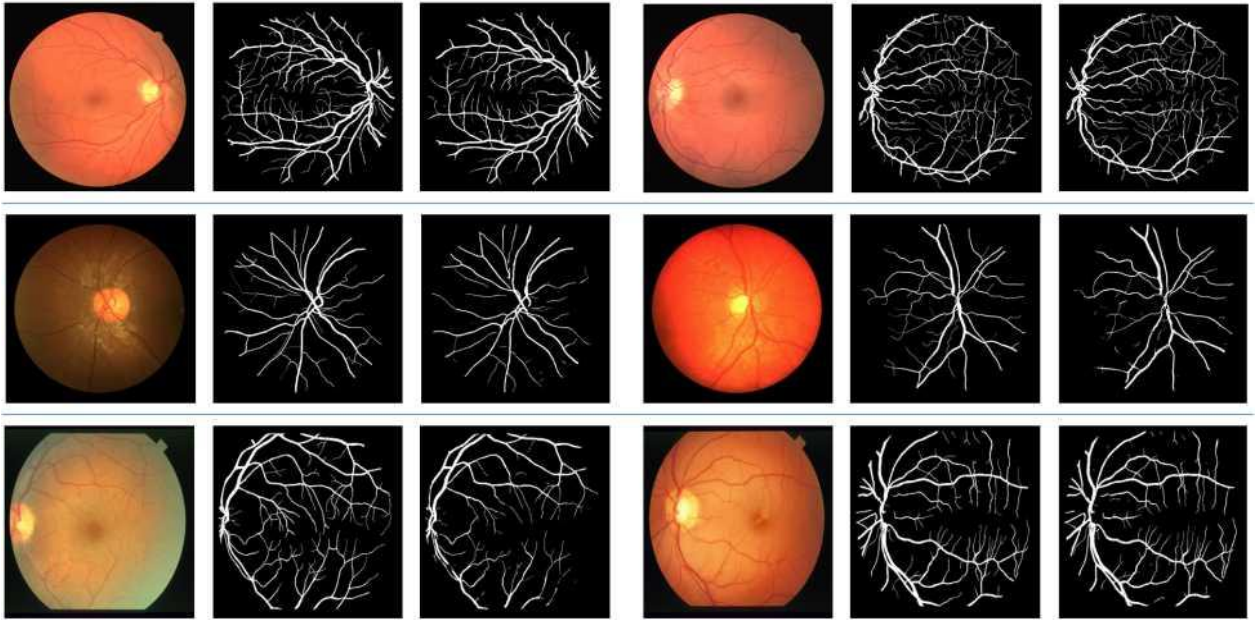

Figure 5: The final results predicted by our ensemble architecture CIEU-Net. We have randomly sampled two color fundus images from each dataset. The three illustrations of each image from left to right are image, ground truth, and prediction. **Top :** Two samples from DRIVE dataset. **Middle :** Two samples from CHASE_DB1 dataset. **Bottom :** Two samples from STARE dataset.

## 5. Experiments design and Results

### 5.1. Implementation details

The experiments have been implemented with the Pytorch framework [51] without using pre-training strategy. During the training process, non-overlapping random cropped image patches have been used as input. Each input patch has been cropped into the size of 64 x 64, and rescale the pixel value to (0,1). Then the Gaussian normalization method have been employed to normalize the input patches with a mean of 0.5 and a standard deviation of 0.5. The batch size has been set to 64, and the number of training steps has been set to 30,000. Cross entropy cost function has been used to evaluate the differences between the output and ground truth and used stochastic gradient descent (SGD) method with Adam optimizer to train our network, whose hyperparameters have been $\beta_1$=0.5 and $\beta_2$=0.999 respectively. And for smoothing of the training process, the L2 regularization has been integrated into the cost function as the weight decay with a weight = 1e-5. Inspired by DeepLab [30], the "poly" learning rate policy has been utilized with the initial learning rate 1e-3. All the matrix calculations have been implemented on 1 NVIDIA TITAN X GPU.

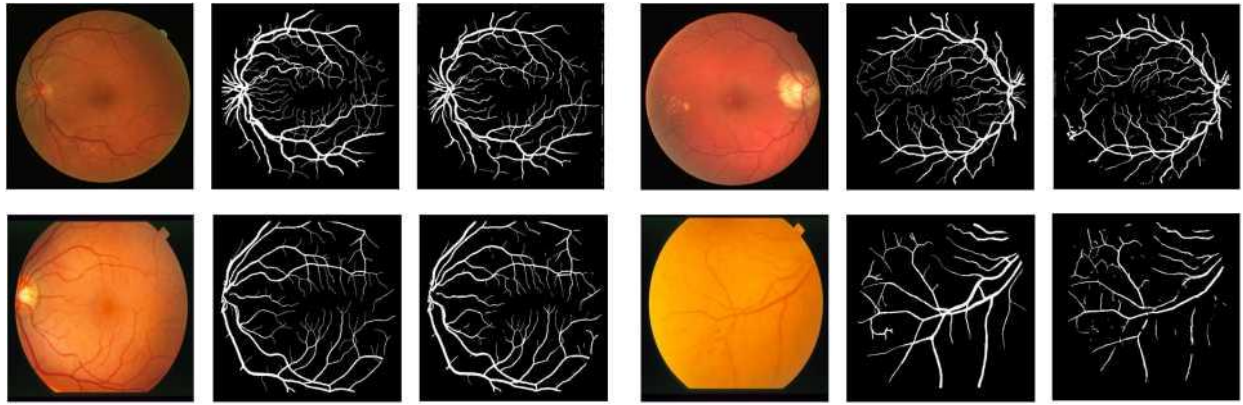

Figure 6: Segmentation results of some abnormal fundus images. The three illustrations of each image from left to right are image, ground truth, and prediction. **Top :** Two samples from DRIVE dataset with abnormality. On the left is Background Diabetic Retinopathy, the right is Pigment Epithelium Changes, Pigmented Scar in Fovea, or Choroidiopathy. **Bottom :** Two samples from STARE dataset with abnormality. On the left is Hypertensive Retinopathy, the right is Hollenhorst Plaque.

In the comparative experiments, the loss curve has shown that the model may have a gradient explode phenomenon. So the L2 norm gradient clipping method has been employed and the threshold has been set to 0.5 to stabilize the training process.

5.2. *Evaluation Metrics*

The purpose of the retinal vessel segmentation task is to classify each pixel in input images into two categories: the blood vessel category (positive), or the background category (negative). By comparing the predicted results of the model with the true values of the ground truth, four types of indicators could be obtained:

1. True Positive (TP): the total number of positive pixels which are correctly predicted;
2. False Positive (FP): the total number of negative pixels which are incorrectly predicted
3. True Negative (TN): the total number of negative pixels which are correctly predicted;
4. False Negative (FN):the total number of positive pixels which are incorrectly predicted;

Through these four basic indicators, some basic evaluation metrics could be calculated such as accuracy (ACC), sensitivity/Recall (SE/Recall), specificity (SP), Precision, and the comprehensive evaluation indicator F1-score. Recall/sensitivity and F1-score are two relatively important metrics in medical image analysis. MCC and AUC are two professional metrics in retinal vessel segmentation. AUC is the area under the Receiver Operating Characteristics curves. In the experiments, nearly all the metrics above mentioned have been used. The calculation formula of the professional metric MCC is shown as follows :

$$MCC \frac{TP/N - S \times P}{\times S \times (1-P) \times (1-S)} \quad (2)$$

where N=TP+TN+FP+FN is the total number of pixels of the image, S=(TP+FN)/N and P=(TP+FP)/N. In addition, the mean intersection of unions (IoU) has also been used which is commonly employed in

$$mIoU \quad \frac{TP}{TP + FP + FN} \quad (3)$$

semantic segmentation tasks.

### 5.3. *Performance on different datasets*

**Results on the DRIVE Dataset.** DRIVE dataset is obtained under the uniform photographic conditions. The proposed CIEU-Net in Fig. 2(b) has been compared with seven state-of-the-art retinal vessel segmentation methods on the DRIVE dataset. Because the source codes of some of the state-of-the-art methods have not been published, the results reported in the original papers have been employed. The results are shown in Table. II. The results show that the proposed CIEU-Net with IN has achieved state-of-the-art performance in sensitivity, F1-score, AUC and MCC on DRIVE dataset.

**Results on the CHASE_DB1 Dataset.** CHASE_DB1 dataset is obtained under the complex photographic conditions. The variances of retina images such as illumination condition and rotation degree could not be uniformly controlled. The proposed CIEU-Net in Fig. 2(b) has been compared with six state-of-the-art retinal vessel segmentation methods on the CHASEDB1 dataset. Same as the experiments on DRIVE dataset, the results reported in the original papers have been employed. The results are shown in Table. III. The results show that the proposed CIEU-Net with BN has achieved state-of-the-art performance in Sensitivity/Recall, F1-score MCC and AUC on CHASEDB1 dataset. However, the results of CIEU-Net with IN is not satisfactory. As a result, for CHASEDB1 dataset, BN should be applied to overcome the intra-dataset variance.

**Results on the STARE Dataset.** STARE dataset is obtained under the complex photographic conditions, which combines healthy and unhealthy data. The proposed CIEU-Net in Fig. 2(b) have been compared with six previous state-of-the-art retinal vessel segmenta-tion methods on the STARE dataset. The results are shown in Table.IV. The results show that our proposed CIEU-Net with BN has achieved state-of-the-art performance.

| Methods | Year | **F1** | **Recall** | **MCC** | ACC | AUC |
|---|---|---|---|---|---|---|
| DeepVessel[24] | 2016 | 0.7542 | 0.7603 | 0.7367 | 0.9523 | - |
| FC-CRF[18] | 2017 | 0.7857 | 0.7897 | 0.7556 | - | - |
| R2U-Net[19] | 2018 | 0.8171 | 0.7792 | - | 0.9556 | - |
| Ladder-Net[20] | 2018 | 0.8202 | 0.7856 | | 0.9663 | - |
| HCF[12] | 2018 | 0.7998 | **0.8176** | 0.7659 | **0.9753** | - |
| Joint-Losses[21] | 2018 | - | 0.7653 | | 0.9542 | 0.9752 |
| M2U-Net[22] | 2019 | - | - | - | 0.9630 | 0.9714 |
| CIEU-Net(BN) | 2019 | 0.8156 | 0.7758 | 0.7673 | 0.9660 | 0.9716 |
| CIEU-Net(IN) | 2019 | **0.8239** | 0.7903 | **0.7722** | 0.9663 | **0.9754** |

Table 2: Results on the DRIVE Dataset. Our CIEU-Net with IN achieves the state-of-the-art performance in F1-score, AUC and MCC.

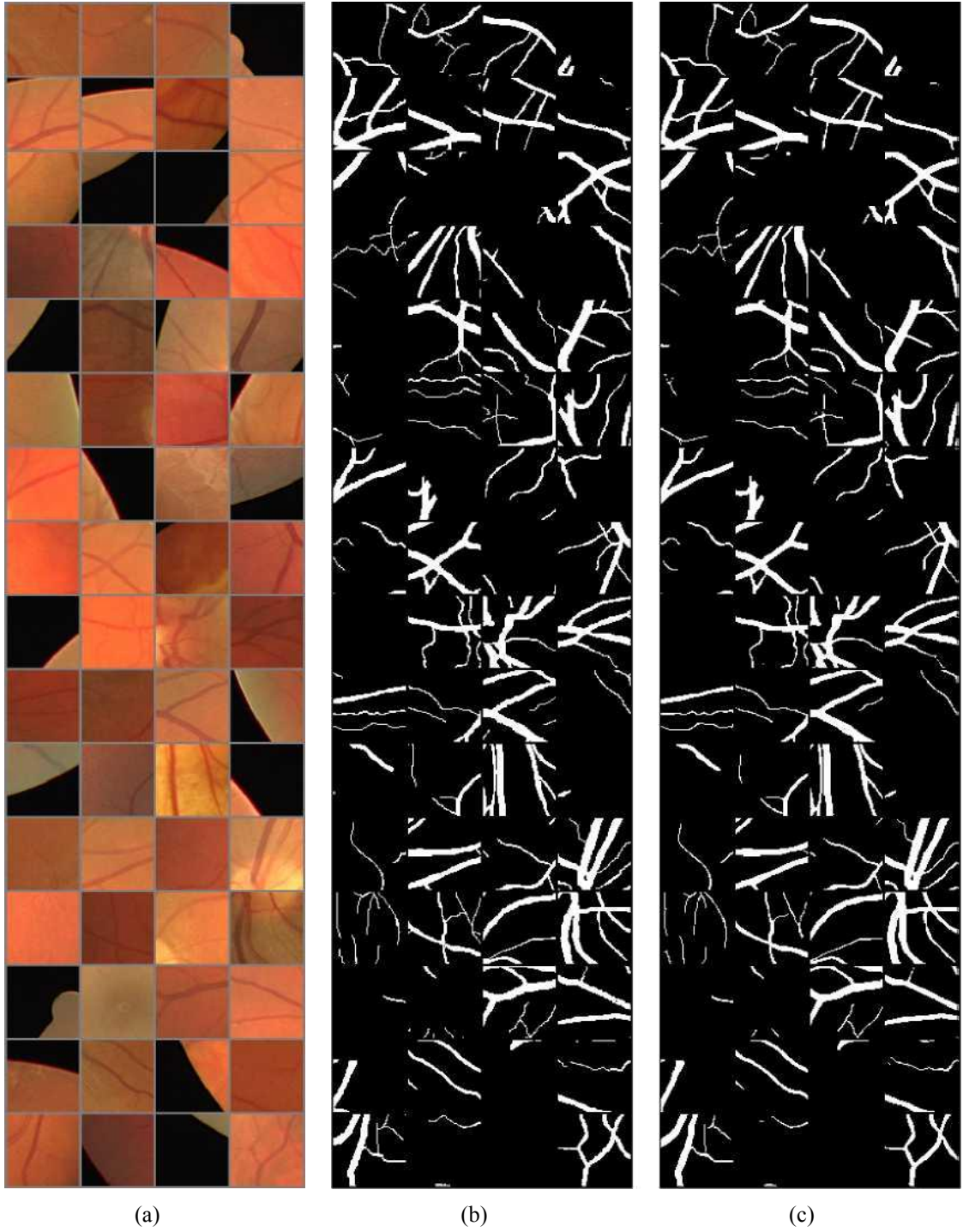

Figure 7: The final results predicted by our ensemble architecture CIEU-Net on patches. The details in the predictions are clear. The predictions are almost the same as the labels. **(a).** Image patches. **(b).** labels. **(c).** predictions.

| Methods | Year | **F1** | **Recall** | **MCC** | ACC | AUC |
|---|---|---|---|---|---|---|
| DeepVessel[24] | 2016 | 0.7304 | 0.7130 | 0.7008 | 0.9486 | - |
| FC-CRF[18] | 2017 | 0.7332 | 0.7277 | 0.7046 | - | - |
| R2U-Net[19] | 2018 | 0.7928 | 0.7978 | - | 0.9634 | - |
| Ladder-Net[20] | 2018 | 0.8031 | 0.7374 | - | 0.9656 | - |
| HCF[12] | 2018 | 0.7646 | 0.7559 | 0.7379 | 0.9518 | - |
| Joint-Losses[21] | 2018 | - | 0.7633 | - | 0.9610 | 0.9781 |
| M2U-Net[22] | 2019 | - | - | - | 0.9703 | 0.9666 |
| CIEU-Net(IN) | 2019 | 0.7639 | 0.7474 | 0.7487 | 0.9648 | 0.9546 |
| CIEU-Net(BN) | 2019 | **0.8045** | **0.7998** | **0.7680** | **0.9746** | **0.9686** |

Table 3: Results on the CHASEDB1 Dataset. Our CIEU-Net with BN achieves the state-of-the-art performance in Recall, F1-score, AUC and MCC.

| Methods | Year | **F1** | **Recall** | **MCC** | ACC | AUC |
|---|---|---|---|---|---|---|
| CrossModality[16] | 2015 | - | 0.7726 | - | 0.9628 | 0.9879 |
| VesselTrack[23] | 2016 | 0.7345 | 0.7527 | 0.7288 | 0.9601 | |
| DeepVessel[24] | 2016 | 0.7304 | 0.7130 | 0.7008 | 0.9486 | - |
| FC-CRF[18] | 2017 | 0.7644 | 0.7680 | 0.7417 | - | - |
| HCF[12] | 2018 | 0.7982 | 0.8239 | 0.7818 | **0.9751** | - |
| Joint-Losses[21] | 2018 | - | 0.7581 | - | 0.9612 | 0.9801 |
| CIEU-Net(IN) | 2019 | 0.8153 | 0.7932 | 0.8001 | 0.9713 | 0.9756 |
| CIEU-Net(BN) | 2019 | **0.8230** | **0.8273** | **0.8075** | 0.9714 | **0.9882** |

Table 4: Results on the STARE Dataset. Our CIEU-Net with BN achieves the state-of-the-art performance in Recall, F1-score, AUC and MCC.

***Illustrations of predictions on these three datasets.*** Two color fundus images have been randomly sampled from each dataset to show our final results. The illustrations are shown in Fig. 5. Fig. 6. illustrates some segmentation results with various abnormalities, with are labeled by dataset annotators. Furthermore, some patches have been randomly sampled to show our performance on details in Fig. 7.

5.4. *Ablation analysis*

In this subsection, sufficient comparisons have been conducted to evaluate the approaches proposed in Section III. The test subset of DRIVE dataset have been chosen for evaluation if there is no additional annotation. In order to verify the generalization of our models, CHASEDB1 and STARE datasets have been also used for evaluation in some specific experiments. In these experiments, more attention has been paid to the evaluation of the network structures, so the metrics more about semantic segmentation has been used.

***Cascaded Dilated Modules.*** Firstly, a U-Net structure baseline model has been built as shown in Fig. 2(a), which has achieved similar performance compared with the U-Net model reproduced by [19]. The baseline model has 35 convolutional layers, 34 normalization layers with 1/16 downsampling stride. The network first encodes and downsamples the input image to 1/16 resolution, then uses skip-connections and layer-by-layer decoder to refine the encoded feature maps to the same resolution of inputs. This structure could be separated into a backbone model of 1/16 strided FCN and a layer-by-layer decoder which could perceive the context information. For verifying the effectiveness of our modules, the performance of cascaded dilated modules have been evaluated on the backbone model of 1/16 strided FCN as shown in Fig. 8.

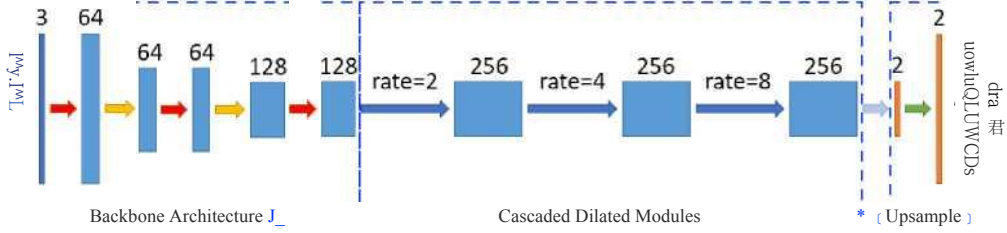

Figure 8: The backbone-based architecture integrated with cascaded dilated modules. The dilated convolutional modules have been stacked for 1, 2 and 3 times after the downsampling progress respectively, making the segmentation resolution 1/16 strided, 1/8strided and 1/4 strided separately.

|  | **F1-score** | **Recall** | SP | ACC | mIoU |
|---|---|---|---|---|---|
| Backbone | 0.0429 | 0.0219 | 0.9988 | 0.9137 | 0.4676 |
| 1/16 Dilated | 0.1033 | 0.0545 | 0.9965 | 0.9144 | 0.4833 |
| 1/8 Dilated | 0.3055 | 0.2184 | 0.9889 | 0.9218 | 0.5579 |
| 1/4 Dilated | 0.6354 | 0.5278 | **0.9866** | 0.9466 | 0.7033 |
| 1/4 Dilated, MG | **0.7161** | **0.6318** | 0.9803 | **0.9500** | **0.7354** |
| U-Net | 0.8090 | 0.7659 | 0.9839 | 0.9658 | 0.8085 |
| U-Net, CDM | **0.8112** | **0.7687** | **0.9851** | **0.9663** | **0.8145** |

Table 5: Experimental results on the backbone-based models. **MG** is the shorthand of Multi-Grid. **CDM** means 3 Cascaded Dilated Modules with the multi-grid strategy. The results show that CDM has shown the effectiveness on the performance with 1/4 strided.

The dilated convolutional modules have been stacked for 1, 2 and 3 times after the downsampling progress respectively, making the segmentation resolution 1/16 strided, 1/8 strided and 1/4 strided separately. Each dilated module has 3 ResBlock-based dilated convolutional blocks. The 1/4 strided dilated model with 3 dilated convolutional blocks occupies similar GPU memory with U-Net baseline model (2873 Mb vs. 2937Mb of baseline). The resolution have not been extended further in consideration of computing cost. As a result, the architecture with 3 dilated convolutional modules (1/4 strided) has shown better performance.

In the 1/4 strided dilated model, the dilation rates of dilated convolutional blocks are {(2,2,2), (4,4,4), (8,8,8)} separately. Following [38], the multi-grid have been applied strategy on each dilated convolutional module, change the dilation rates of dilated convolutional blocks to {(1,2,1), (2,4,2), (4,8,4)}. In Fig. 4, the illustrations show that multi-grid strategy provides continuous receptive fields, which is beneficial for vessel segmentation. The cascaded dilated modules have been also tested on U-Net baseline model (adding layer-by-layer decoder) to examine the capacity of cascaded dilated modules.

The results of models with cascaded dilated modules are listed in Table. V. It is clear that dilated modules could obviously improve the performance of vessel segmentation. The cascaded dilated modules could preserve the resolution of features and avoid the loss of details caused by downsampling operations. Multi-grid strategy has been proved to be essential for cascaded dilated modules in view of the significant performance boost. Moreover, the 1/16 downsampled model has degenerated seriously in segmentation accuracy without layer-bylayer decoder. Different from our examinations, models designed for semantic segmentation task with 1/4 downsampling operation would not result in large accuracy degeneration compared with U-Net structure models, because the domain-specific performance bottleneck in semantic segmentation task is generally the complexity of variant classes and objects in scenes. The experimental results have proved that the key points in the vessel segmentation task are details and edges so that the decoder component should be employed in vessel segmentation models. The experimental results have revealed the joint effects of layer-by-layer decoder and cascaded dilated modules.

*Contextual Information Enhancing Module.* The effectiveness of the Contextual Information Enhancing modules has been further tested on the backbone-based models. The Contextual Information Enhancing modules have been attached after the cascaded dilated ResBlock modules to generate segmentation predictive masks. The results are listed in Table. VI. It is clear that the models could largely benefit from Contextual Information Enhancing modules, even on 1/16 and 1/4 downsampled resolution. PSP module could generate more aggregated and continuous pixel patches. On the contrary, ASPP module is able to predict some incontinuous vessel regions. Visualizations of results about Contextual Information Enhancing modules have been shown in Fig. 9. In the backbone-based models under 1/4 downsampling without layer-by-layer decoder, the results show that with more incontinuous pixels, ASPP has shown better performance, as shown in Fig. 9, Line 1. However, in the U-Net based models with layer-by-layer decoder, tiny vessels/branch vessels have been preferably predicted. In this situation, these two contextual modules have been tested on the ensemble model separately, as described in Table. VI. The results show that these improperly predicted incontinuous pixels generated by ASPP module are disadvantageous for continuous vessel prediction, as shown in Fig. 9, Line 2. At last, the PSP module have been chosen as the contextual module in our ensemble architecture for retinal vessel segmentation.

*Normalization options.* Different normalization options of the network have been tested, as proposed in Section III. Firstly, these approaches have been examined in the U-

|                     | **F1 score** | **Recall** | SP     | ACC    | **mIoU** |
|---------------------|--------------|------------|--------|--------|----------|
| Backbone            | 0.0429       | 0.0219     | 0.9988 | 0.9137 | 0.4676   |
| Backbone, ASPP      | 0.0533       | 0.0274     | 0.9984 | 0.9139 | 0.4703   |
| Backbone, PSP       | 0.0473       | 0.0242     | 0.9987 | 0.9138 | 0.4687   |
| 1/4 dilated         | 0.6354       | 0.5278     | **0.9866** | 0.9466 | 0.7033 |
| 1/4 dilated, ASPP   | **0.7050**   | **0.6165** | 0.9816 | **0.9498** | **0.7318** |
| 1/4 dilated, PSP    | 0.6895       | 0.5956     | 0.9831 | 0.9494 | 0.7264   |
| U-Net, CDM          | 0.8112       | 0.7687     | 0.9851 | 0.9663 | 0.8145   |
| U-Net, CDM, ASPP    | 0.8035       | 0.7564     | **0.9854** | 0.9654 | 0.8094 |
| U-Net, CDM, PSP     | **0.8156**   | **0.7758** | 0.9848 | **0.9666** | **0.8166** |

Table 6: Ablation studies for Contextual Information Enhancing modules. CDM means the U-Net is modified by 3 cascaded dilated modules with multi-grid strategy. The last line is our ensemble model. The results show that those Contextual Information Enhancing modules have shown their effectiveness on the performance.

| DRIVE         | **F1-score** | **Recall** | SP         | ACC        | **mIoU**   |
|---------------|--------------|------------|------------|------------|------------|
| U-Net, BN     | 0.8090       | 0.7659     | 0.9839     | 0.9658     | 0.8085     |
| U-Net, GN     | 0.8152       | 0.7753     | 0.9846     | **0.9664** | 0.8157     |
| U-Net, IN     | **0.8213**   | **0.7859** | 0.9833     | 0.9661     | **0.8162** |
| Ensemble, BN  | 0.8156       | 0.7758     | **0.9848** | 0.9660     | 0.8166     |
| Ensemble, IN  | **0.8239**   | **0.7903** | 0.9830     | **0.9663** | **0.8174** |
| CHASEDB1      |              |            |            |            |            |
| Ensemble, BN  | **0.8045**   | **0.7998** | **0.9857** | **0.9746** | **0.8073** |
| Ensemble, IN  | 0.7639       | 0.7374     | 0.9787     | 0.9648     | 0.7546     |
| STARE         |              |            |            |            |            |
| Ensemble, BN  | **0.8230**   | **0.8273** | **0.8075** | **0.9714** | **0.9882** |
| Ensemble, IN  | 0.8153       | 0.7932     | 0.8001     | 0.9713     | 0.9756     |

Table 7: Results of Different normalization options. As a result, different normalization methods should be chosen properly based on the properties of specific retinal datasets. In DRIVE, IN is the better choice. In CHASEDB1 and STARE, BN is the better choice.

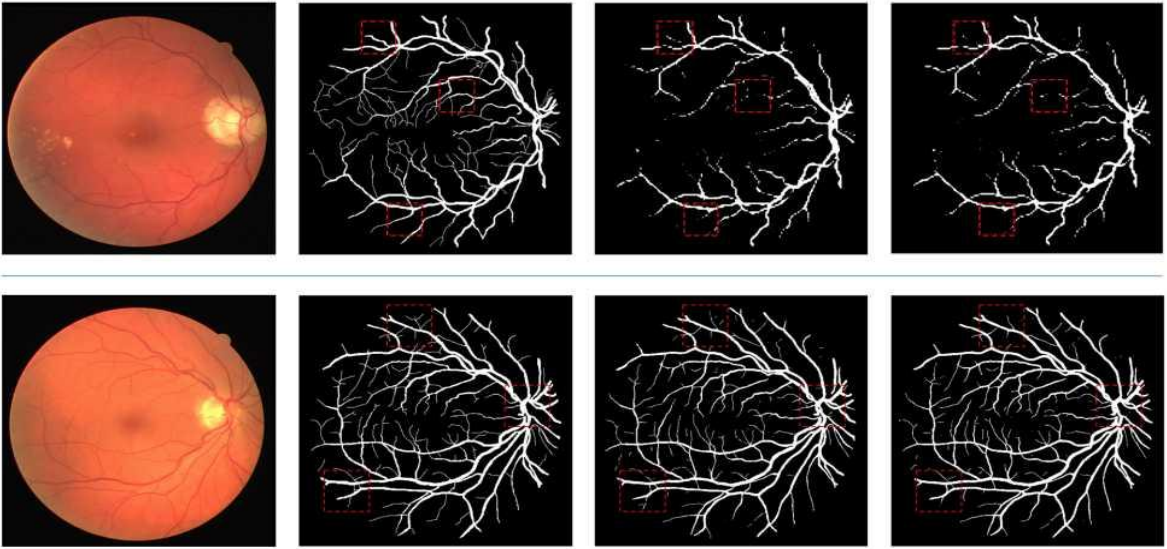

Figure 9: The illustrations of comparisons between PSP and ASPP modules. There are two examples in two lines. In each line, there are four illustrations : image, label, predition of ASPP, and prediction of PSP. **Line 1 :** The prediction of backbone-based models under 1/4 downsampling without layer-by-layer decoder. ASPP has predicted more vessel regions than PSP module as shown inside red dashed border. **Line 2 :** The prediction of U-Net based models under 1/4 downsampling with layer-by-layer decoder. The incontinuous pixel regions are redundant and sometimes harmful for accurate vessel prediction on the high-resolution segmentation mask. The PSP module with spatial-continuity has a better match with layer-by-layer decoder than ASPP.

Net baseline models on DRIVE dataset. Normalization options have also been compared on our final ensemble model for DRIVE and CHASEDB1 dataset to verify our hypothesis. From Table. VII, the indicators show that models with IN show the best performance on DRIVE dataset. GN is abnormally better than BN. The final ensemble model on DRIVE dataset supports this conclusion because the IN branch also achieves the best result. Just the opposite, BN branch shows higher segmentation accuracy on CHASE_DB1 dataset which is more variant than DRIVE dataset. These experimental results further improve our knowledge of FCN-based segmentation methods with respect to retinal vessel segmentation task. The normalization methods should be chosen properly based on the properties of specific retinal datasets. This strategy about normalization is potential to extend into other tasks of medical image analysis.

*Modules ablation study.* The effectiveness of each proposed module in section III have been compared, the results are shown in Table. VIII. For an intuitive understanding, one sample have been illustrated to show the improvement of different modules, as shown in Fig. 10. The cascaded dilated convolutional modules have been integrated into the encoderdecoder structure under 1/4 resolution, as shown in Fig. 2(b). It is clear that cascaded dilated modules, PSP module, and instance normalization could boost the segmentation

| Decoder | CD | PSP | BN | IN | F1-score | Recall | SP | ACC | mIoU |
|---------|----|----|----|----|----------|--------|-----|------|------|
|         |    |    | /  |    | 0.0429   | 0.0219 | 0.9988 | 0.9137 | 0.4676 |
| /       |    |    | /  |    | 0.8090   | 0.7659 | 0.9839 | 0.9658 | 0.8085 |
| /       | /  |    | /  |    | 0.8112   | 0.7687 | 0.9851 | 0.9663 | 0.8145 |
| /       | /  | /  | /  |    | 0.8156   | 0.7758 | 0.9848 | 0.9660 | 0.8166 |
| /       | /  | /  |    | /  | **0.8239** | **0.7903** | **0.9830** | **0.9663** | **0.8174** |

Table 8: Ablation studies for modules. The first line is the backbone model of 1/16 strided FCN. The second line is the U-Net baseline model. When applying IN, all batch normalization layers but those in context fusion module are replaced

by instance normalization. The final line is our proposed CIEU-Net with IN. Our CIEU-Net shows the best performances on all evaluation metrics. (CD: Cascaded Dilated, PSP: PSP Module) accuracy. The ensemble model with cascaded dilated modules, PSP module, and IN has achieved the state-of-the-art performance on DRIVE dataset. However, applying the ASPP module results in lower accuracy. We have shown visualized masks of ASPP module and PSP module in Fig. 10. From the visualization, the results demonstrate that PSP module can produce more aggregated vessel regions with better shape. Compared with PSP module, the result of ASPP module includes many small dissociated regions which do not appear in vessel label regions, due to the discontinuous dilated convolutions. We believe that this is the reason why accuracy become degraded after applying ASPP module. In our final ensemble model for evaluation, the PSP module have been employed with spatial-continuity property for Contextual Information Enhancement.

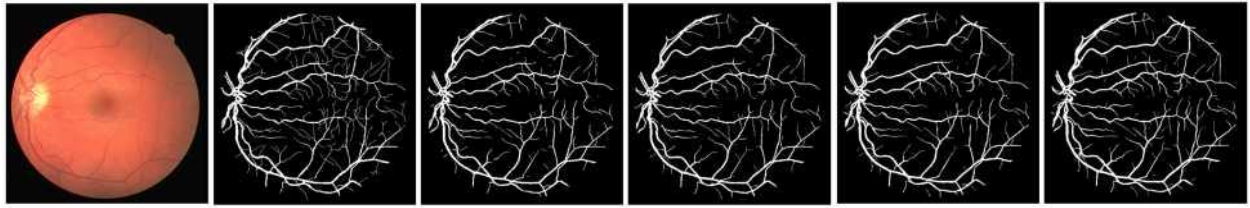

Figure 10: The illustrations of the improvements of different modules. From left to right: (1). Original image, (2). Label, (3). Result of U-net, (4). Result of Unet + CDM, (5). Result of Unet + CDM + Contextual module, (6). Result of CIEU-net. The result gets better in details with the increase of modules.

## 6. Limitation

In our study, a CIEU-Net have been proposed for retinal vessel segmentation. Fortunately, our method has achieved satisfactory results. However, there are also some limitations to our research. Firstly, more attention have been paid to the final performance. the calculating cost of our network have been ignored . Compared with the baseline model U-Net, our architecture has taken 1.5 times longer. Though only 1 GPU have been used for calculation. For embedding into the computer system, our model is suitable. But in the application of portable equipment, this might not be lightweight enough. And this phenomenon has limited our application scenarios. Secondly, the datasets selected are circumscribed. Although there are some unhealthy images, these datasets might not include holistic pathological morphology. The capabilities of our model are limited in terms of robustness.

## 7. Conclusion

In this paper, a deep learning based architecture for retinal vessel segmentation have been proposed which demonstrates the advantage of jointly utilizing semantic segmentation modules and the basic medical image segmentation methods. Extensive comparative evaluations on color fundus image datasets demonstrate that the proposed method is more accurate and efficient. In the future, more attention should be paid to explore the more lightweight and efficient architectures for the embedding on medical devices, promoting the process from research to engineering.

## 8. Acknowledgment

Thanks to the creators of these three datasets DRIVE[52], CHASEDB1[53], and STARE[54] for their contribution to the segmentation of the fundus vessels.